\documentclass[final,1p,times]{elsarticle}

\usepackage{amssymb}
\usepackage{amsthm}
\usepackage{amsmath}
\usepackage{graphicx}
\usepackage{hyperref}

\usepackage{maplestd2e} %fuer .mw

\newcounter{bla}

\journal{Computer Physics Communications}

\newcommand{\R}{\mathbb R}
\newcommand{\Q}{\mathbb Q}
\newcommand{\C}{\mathbb C}
\newcommand{\Z}{\mathbb Z}

\newcommand{\ld}{{\rm ld}}
\newcommand{\initial}{{\rm init}}
\newcommand{\separant}{{\rm sep}}

\theoremstyle{definition}
\newtheorem{definition}{Definition}
\newtheorem{remark}[definition]{Remark}
\newtheorem{example}[definition]{Example}

\theoremstyle{plain}
\newtheorem{theorem}[definition]{Theorem}
\newtheorem{cor}[definition]{Corollary}

\begin{document}

\begin{frontmatter}



\title{The MAPLE package TDDS for computing Thomas decompositions of systems of nonlinear PDEs}


\author[a,b]{Vladimir P. Gerdt}
\author[c]{Markus Lange-Hegermann}
\author[d]{Daniel Robertz\corref{author}}

\cortext[author] {Corresponding author.\\\textit{E-mail address:} daniel.robertz@plymouth.ac.uk}
\address[a]{Laboratory of Information Technologies, Joint Institute for Nuclear Research, 6~Joliot-Curie St, Dubna, 141980, Russian Federation}
\address[b]{Russian Federation and Peoples' Friendship University of Russia (RUDN University), 6~Miklukho-Maklaya St, Moscow, 117198, Russian Federation}
\address[c]{Department of Electrical Engineering and Computer Science, Ostwestfalen-Lippe University of Applied Sciences, Liebigstra{\ss}e 87, 32657 Lemgo, Germany}
\address[d]{School of Computing, Electronics and Mathematics, Plymouth University, 2-5~Kirkby~Place, Drake Circus, Plymouth, PL4~8AA, United Kingdom}

\begin{abstract}
We present the Maple package TDDS (Thomas Decomposition of Differential Systems).
Given a polynomially nonlinear differential system, which in addition to equations may contain inequations, this package computes a decomposition of it into a finite set of differentially triangular and algebraically simple subsystems whose subsets of equations are involutive.
Usually the decomposed system is substantially easier to investigate and solve both analytically and numerically. The distinctive property of a Thomas decomposition is disjointness of the solution sets of the output subsystems. Thereby, a solution of a well-posed initial problem belongs to one and only one output subsystem. The Thomas decomposition is fully algorithmic. It allows to perform important elements of algebraic analysis of an input differential system such as: verifying consistency, i.e., the existence of solutions; detecting the arbitrariness in the general analytic solution; given an additional equation, checking whether this equation is satisfied by all common solutions of the input system; eliminating a part of dependent variables from the system if such elimination is possible; revealing hidden constraints on dependent variables, etc. Examples illustrating the use of the package are given.
\end{abstract}


\end{frontmatter}

% \linenumbers

% Computer program descriptions should contain the following
% PROGRAM SUMMARY.

{\bf PROGRAM SUMMARY}
  %Delete as appropriate.

\begin{small}
\noindent
{\em Program Title:} TDDS\\
{\em Licensing provisions:} GNU LPGL license\\
%CC0 1.0/CC By 4.0/MIT/Apache-2.0/BSD 3-clause/BSD 2-clause/GPLv3/CC BY NC 3.0 }                                   \\ 
%Markus: I prefer the CPC licence http://cpc.cs.qub.ac.uk/licence/licence.html
{\em Programming language:} MAPLE~11 to MAPLE~2017, available independently in MAPLE~2018\\
%
%{\em Supplementary material:}                                 \\
  % Fill in if necessary, otherwise leave out.
%{\em Journal reference of previous version:}                  \\
  %Only required for a New Version summary, otherwise leave out.
%{\em Does the new version supersede the previous version?:}   \\
  %Only required for a New Version summary, otherwise leave out.
%{\em Reasons for the new version:}\\
  %Only required for a New Version summary, otherwise leave out.
%{\em Summary of revisions:}*\\
  %Only required for a New Version summary, otherwise leave out.
%
{\em Nature of problem(approx. 50-250 words):}\\
Systems of polynomially nonlinear partial differential equations
are not given in a formally integrable form in general. In order
to determine analytic solutions in terms of power series,
symbolic manipulations are necessary to find a complete set of
conditions for the unknown Taylor coefficients. A particular case
of that problem is deciding consistency of a system of PDEs.
Nonlinear PDEs require splitting into different cases in general.
Deciding whether another PDE is a consequence of a given system
depends on similar symbolic manipulations. Computing all consequences of a given system which involve only a subset of the unknown functions or a certain subset of their derivatives are instances of differential elimination problems, which arise, e.g., in detection of hidden constraints in singular dynamical systems and field theoretical models.\\
{\em Solution method(approx. 50-250 words):}\\
The solution method consists, in principle, of
pseudo-division of differential polynomials, as in
Euclid's algorithm, with case distinctions according
to vanishing or non-vanishing leading coefficients
and discriminants, combined with completion to
involution for partial differential equations.
Since an enormous growth of expressions can be
expected in general, efficient versions of these
techniques need to be used, e.g., subresultants,
Janet division, and need to be applied in an
appropriate order. Factorization of polynomials,
while not strictly necessary for the method,
should be utilized to reduce the size of expressions
whenever possible.
%{\em Additional comments including Restrictions and Unusual features (approx. 50-250 words):}\\
  %Provide any additional comments here.
   \\

%\begin{thebibliography}{0}
%\bibitem{1}Reference 1         % This list should only contain those items referenced in the                 
%\bibitem{2}Reference 2         % Program Summary section.   
%\bibitem{3}Reference 3         % Type references in text as [1], [2], etc.
%                               % This list is different from the bibliography at the end of 
%                               % the Long Write-Up.
%\end{thebibliography}
%* Items marked with an asterisk are only required for new versions
%of programs previously published in the CPC Program Library.\\
\end{small}

\section{Introduction}
\label{sec:intro}

The Maple package TDDS (Thomas Decomposition of Differential Systems) is applicable to a set of finite-order partial differential equations (PDEs) of the form (cf. Section~\ref{sec:theory})
\begin{equation}\label{pde}
p_i\left(x_{1},\ldots,x_n;u_1,\ldots,u_m,\ldots,\frac{\partial^{j_1+\cdots +j_n} u_k}{\partial x_1^{j_1}\cdots \partial x_n^{j_n}},\ldots\right)=0\,,\quad i=1,\ldots, s\,,
\end{equation}
where $k=1,\ldots,m$, $u_k=u_k(x_1,\ldots,x_n)$. It is assumed that the left hand sides $p_i$ in~\eqref{pde} are polynomials in their arguments. The package also allows enlargement of (1) with a set of inequations $\{q_1\neq 0,\ldots,q_t\neq 0\}$ where $q_k$ $(k=1,\ldots,t)$ are also polynomials in the independent variables $x_1,\ldots,x_n$, dependent variables $u_1,\ldots,u_m$ and their partial derivatives. 

A constructive algebraic approach to study systems of the form \eqref{pde} goes back to the following classical theorem proved by Kovalevskaya~\cite{Kovalevskaya} (cf.~\cite{Petrovsky}). 

\begin{theorem}[Cauchy-Kovalevskaya theorem]\label{thm:cauchy-kovalevskaya}
Let the left hand sides in system~\eqref{pde} read 
\begin{equation}\label{KovalevskayaForm} 
p_i \, = \,
\frac{\partial^{n_i} u_i}{\partial x_1^{n_i}}-F_i\left(x_1,\ldots,x_n;u_1,\ldots,u_m,\ldots,\frac{\partial^{j_1+\cdots +j_n} u_k}{\partial x_1^{j_1}\cdots \partial x_n^{j_n}},\ldots\right)\,, \qquad i=1,\ldots,m=s.
\end{equation}
where $j_1+\cdots+j_n\leq n_i\,,\ j_1<n_i$ and all the functions $F_i$ (not necessarily polynomial) are analytic in a neighborhood of the point
\begin{equation}\label{Cauchy-Kovalevskaya_IC}
x_i=x_i^0,\quad u_k=u_k^0,\quad
\left. r^0_{k;\,j_1,\ldots,j_n}:=\frac{\partial^{j_1+\cdots +j_n} u_k}{\partial x_1^{j_1}\cdots \partial x_n^{j_n}}\right|_{x_1=x^0_1, \ldots, x_n=x^0_n} \qquad (i,k=1,\ldots,m=s)\,.
\end{equation}
Then in some neighborhood of the point $(x_1^0, \ldots, x_n^0)$ the PDE system $p_i=0$ $(i=1,\ldots,s)$ has a unique analytic solution satisfying the initial conditions
\begin{equation}\label{IC}
\left\{
\begin{array}{l}
u_k=\phi_k(x_2,x_3,\ldots,x_n)\,,\\[0.2em]
\frac{\partial u_k}{\partial x_1} = \phi_{k;\,1}(x_2,x_3,\ldots,x_n)\,,\\[0.2em]
..........................................\\[0.2em] 
\frac{\partial^{n_k-1}u_k}{\partial x_1^{n_k-1}}=\phi_{k;\,n_k-1}(x_2,x_3,\ldots,x_n)\,,
\end{array}
\right. \text{for}\ x_1=x_1^0\,,\ k=1,\ldots,m\,,\\[0.2em]
\end{equation}
where all $\phi$ are arbitrary analytic functions of their arguments in a neighborhood of the point $(x_2^0,\ldots,x_n^0)$ such that they take at this point the initial values satisfying \eqref{Cauchy-Kovalevskaya_IC} and~\eqref{IC}.
\end{theorem}

Riquier~\cite{Riquier}, Janet~\cite{Janet1} and Thomas~\cite{ThomasRiquier1,Thomas} developed a framework for generalization of Theorem~\ref{thm:cauchy-kovalevskaya}. Riquier introduced a ranking $\succ$ on partial derivatives (see Section~\ref{sec:theory},~Definition~\ref{de:ranking}), called {\em Riquier ranking} (cf.\ Definition~\ref{def_riquier} or \cite{RustReid}).
Given a Riquier ranking $\succ$, a PDE system~\eqref{pde} is {\em orthonomic}~\cite{Riquier} if each of its equations is solved with respect to the highest ranked partial derivative occurring in the equation, and hence has the form
\begin{equation}\label{orthonomic}
p_i=0\,,\quad p_i:=\delta_i u_{k_i}-F_i\,,\quad i \in \{ 1, \ldots, s \}\,,\quad k_i \in \{ 1, \ldots, m \}\,,
\end{equation}
where the highest ranked derivative in $p_i$, i.e.\ its {\em leader} (see Section~\ref{sec:theory}), is written as $\delta_i u_{k_i}$ with a differential operator $\delta_i$. This derivative is called {\em principal} and the derivatives occurring in $F_i$ are called {\em parametric}. The orthonomic system~\eqref{orthonomic} is called {\em passive} if its differential and algebraic consequences do not lead to additional constraints on the parametric derivatives. It should be noted that a PDE system in the Kovalevskaya form~\eqref{KovalevskayaForm} is orthonomic and passive for a certain Riquier ranking~\cite{ThomasRiquier1}.

Riquier \cite{Riquier} proved the existence of analytic solutions for orthonomic and passive systems of PDEs. Janet~\cite{Janet1} designed algebraic criteria of passivity for orthonomic systems in terms of monomials associated with the principal derivatives in accordance to the mapping
\begin{equation}\label{association}
\frac{\partial^{j_1+\cdots +j_n}}{\partial x_1^{j_1}\cdots \partial x_n^{j_n}} \, \longmapsto \, x_1^{j_1}\cdots x_n^{j_n}\,.
\end{equation}
For $k \in \{1,2,\ldots,m\}$ let $M_k$ be the {\em leading monomial set}, the set of monomials associated by~\eqref{association} with the principal partial derivatives of $u_k$. The Janet criteria for passivity are based on a certain partition of variables for each $w\in M_k$:
\begin{equation*}\label{separation}
\{x_1,\ldots,x_n\}={\cal{M}}(w,M_k)\cup {\cal{NM}}(w,M_k)\,.
\end{equation*} 
Moreover, for a linear PDE system the algebraic criteria allow to transform it algorithmically into a passive form. 

By using the results of Riquier and Janet, Thomas~\cite{ThomasRiquier1} formulated a Cauchy problem providing the uniqueness and existence of an analytic solution in terms of the monomials associated with the parametric derivatives (complementary monomials). Furthermore, Thomas~\cite{Thomas,ThomasRoots} generalized the ideas and methods of Riquier-Janet theory~\cite{Riquier,Janet1} to PDE systems of form~\eqref{pde}. He showed that such differential systems can be decomposed into finitely many passive subsystems. In so doing, each of these subsystems which we call {\em simple differential systems}\footnote{Thomas in~\cite{Thomas} called them {\em passive standard systems}.} has certain triangular structure and can be solved with respect to its leaders such that the solved system is passive and orthonomic. It admits~\cite{Thomas,ThomasRoots} posing of a Cauchy problem with initial data generalizing those in~\eqref{Cauchy-Kovalevskaya_IC},~\eqref{IC} and providing the uniqueness and (for a Riquier ranking) existence of an analytic solution. 
Based on the ideas of Janet and Thomas, the foundations of differential algebra were developed by Ritt \cite{Ritt}. Then Wu \cite{WuDifferentialAlgebraicGeometry} (cf.\ also \cite{LiWang}) further developed the characteristic set method introduced by Ritt.
The first implementation of Thomas decomposition for systems of algebraic equations and ordinary differential systems was developed by Wang
\cite{WangEliminationMethods,WangEliminationPractice}.

In our papers~\cite{GerdtThomas,BaechlerGerdtLangeHegermannRobertz2} (see also the book~\cite{Robertz6}) the Thomas approach was algorithmized and implemented in full generality in Maple. It should be emphasized that the Thomas decomposition is different from two other methods of decomposition into triangular and passive differential subsystems based on Ro\-sen\-feld-Gr\"{o}b\-ner~\cite{BoulierLazardOllivierPetitotAAECC} and rif~\cite{ReidWittkopfBoulton} algorithms\footnote{Implementations of the Ro\-sen\-feld-Gr\"{o}b\-ner and rif algorithms have been available as packages in the standard Maple distribution. The package discussed in this paper has now also been made available in Maple 2018.}, respectively. In distinction to those methods the Thomas decomposition method combines {\em disjointness} of the solution sets of the output subsystems with the decomposition into {\em characterizable} differential ideals \cite{HubertSurvey}.
These properties are not obtained by a Gr\"{o}bner basis of the initial differential ideal, e.g. the basis introduced by Mansfield~\cite{MansfieldThesis}.
A related difficult problem is to decide to which prime component of a radical differential ideal a given solution belongs, where important contributions have been made, e.g., in \cite{RittSingularSolutions}, \cite{HubertEssentialComponents}.
In addition, the strategy for completing differential systems to passive ones in a Thomas decomposition is based on Janet's criteria mentioned above.
For a more detailed comparison we refer to \cite[Subsect.~4.5]{BaechlerGerdtLangeHegermannRobertz2}.

The present paper is organized as follows. In Section~\ref{sec:thomas} we illustrate by an explicit example of nonlinear PDE systems some features of the differential Thomas decomposition. Basic theoretical aspects of simple differential systems are described in Section~\ref{sec:theory} including   
underlying definitions and statements. Section~\ref{sec:commands} presents a list of commands of the package TDDS, and the most important ones of them are illustrated by examples in Section~\ref{sec:examples}. 

\section{Thomas decomposition of nonlinear PDE systems}
\label{sec:thomas}

Among the features of a Thomas decomposition of a system of nonlinear PDEs
is the possibility to determine power series solutions of the system
around a sufficiently generic point
in a straightforward way, to decide whether another PDE is a consequence
of the system, and to solve differential elimination problems. We illustrate
these features with explicit examples.

As a first example we consider the following system of nonlinear PDEs
for one unknown function $u(t, x)$
\begin{equation}\label{eq:KdV}
\left\{
\begin{array}{rcl}
u_t - 6 \, u \, u_x + u_{x,x,x} & = & 0\,,\\[0.2em]
u \, u_{t,x} - u_t \, u_x & = & 0\,,
\end{array}
\right.
\end{equation}
which is a combination of the Korteweg-de Vries equation
and a Wronskian determinant expressing that $u(t, x)$ is
a product of a function of $t$ and a function of $x$
(cf.\ also \cite[Ex.~2.2.61]{Robertz6}).
We would like to determine all power series solutions (around $(0, 0)$)
\begin{equation}\label{eq:Taylor}
u(t, x) \, \, = \, \, \sum_{j=0}^{\infty} \sum_{k=0}^{\infty} a_{j,k} \,
\frac{t^j}{j!} \, \frac{x^k}{k!}\,,
\qquad a_{j,k} \in \C\,,
\end{equation}
of system (\ref{eq:KdV}).
The idea is to partition the set of unknown coefficients $a_{j,k}$ into
two subsets: Taylor coefficients whose values can be chosen arbitrarily
(up to certain genericity assumptions, cf. below), and Taylor coefficients
whose values are determined from the chosen values (as roots of certain
univariate polynomials in general) by substituting (\ref{eq:Taylor}) into
the system of PDEs and comparing coefficients.
However, this procedure requires preparatory symbolic manipulations of (\ref{eq:KdV})
in order to ensure \emph{passivity} (or \emph{formal integrability}), i.e.\ to ensure that comparing
coefficients results in a complete set of conditions for $a_{j,k}$.

The Thomas decomposition method transforms any given system of polynomially
nonlinear partial differential equations (and/or inequations) into an equivalent
finite collection of so-called simple differential systems. The solution sets
of the resulting simple differential systems form a partition of the solution
set of the original PDE system. Moreover, each simple differential system
is formally integrable and allows to solve for the coefficients $a_{j,k}$
in (\ref{eq:Taylor}) successively.

Different strategies, e.g., concerning the order in which partial derivatives
of $u(t, x)$ and the corresponding Taylor coefficients $a_{j,k}$ are dealt
with are possible.
More precisely, the left hand sides of equations (and inequations) can be
expressed as univariate polynomials in their highest ranked partial derivatives
with coefficients that are polynomials in lower ranked partial derivatives,
which can be recursively represented in the same way.
The Thomas decomposition method depends on a given \emph{ranking}, i.e.,
a total order $\succ$ on the set of all partial derivatives
\[
\left\{ \, \frac{\partial^{j+k} u}{\partial t^j \, \partial x^k} \; \middle| \;
j, k \in \Z_{\ge 0} \, \right\}
\]
which respects differentiation, i.e., for all $j_1$, $k_1$, $j_2$, $k_2$, $m$, $n \in \Z_{\ge 0}$
we have
\[
\frac{\partial^{j_1+k_1} u}{\partial t^{j_1} \, \partial x^{k_1}} \succ
\frac{\partial^{j_2+k_2} u}{\partial t^{j_2} \, \partial x^{k_2}}
\qquad \Longrightarrow \qquad
\frac{\partial^{j_1+m+k_1+n} u}{\partial t^{j_1+m} \, \partial x^{k_1+n}} \succ
\frac{\partial^{j_2+m+k_2+n} u}{\partial t^{j_2+m} \, \partial x^{k_2+n}}\,,
\]
and which does not admit infinitely descending chains, i.e.,
\[
\frac{\partial^{j+k} u}{\partial t^j \, \partial x^k} \succ u
\qquad \mbox{for all } j, k \in \Z_{\ge 0}\,, \quad j+k > 0\,.
\]
For the given example we choose the ranking $\succ$ which is defined by
\begin{align}\label{eq_ranking}
\frac{\partial^{j_1+k_1} u}{\partial t^{j_1} \, \partial x^{k_1}} \succ
\frac{\partial^{j_2+k_2} u}{\partial t^{j_2} \, \partial x^{k_2}}
\quad \Longleftrightarrow \quad
\left\{
\begin{array}{l}
j_1 + k_1 > j_2 + k_2 \qquad \mbox{or}\\[0.2em]
(\, j_1 + k_1 = j_2 + k_2 \quad \mbox{and}
\ (\, k_1 < k_2\\[0.2em]
\ \mbox{or} \quad k_1 = k_2 \quad \mbox{and} \quad j_1 < j_2\, )),
\end{array}
\right.
\end{align}
which is analogous to the degree-reverse lexicographical term ordering used
for Gr{\"o}bner basis computations \cite{Sturmfels}.
More generally, when dealing with $n$
independent variables, the \emph{degree-reverse lexicographical ranking} is
defined by comparing the total differentiation order and in case of equality
defining the greater derivative to be the one involving least differentiation
when comparing lexicographically from $\partial_n$ backwards to $\partial_1$.

Given a system of polynomially nonlinear partial differential equations
(and/or inequations) and a ranking $\succ$, the Thomas decomposition method
computes linear combinations of partial derivatives of equations so as to
eliminate highly ranked partial derivatives as much as possible. The
polynomial reductions which are applied for this purpose assume that the
coefficient of the highest power of the highest ranked derivative in the
divisor does not vanish on the solution set of the system. In general this
requires case distinctions. The Thomas decomposition method splits systems
by introducing new equations and the complementary inequations
as necessary for performing reductions and ensuring the equivalence of
the resulting decomposition to the original system. Hence, one may think
of the construction of a Thomas decomposition as a binary tree whose leaves
correspond to the resulting simple differential systems.

For system (\ref{eq:KdV}) the resulting Thomas decomposition (with respect
to $\succ$) consists of three simple differential systems:
\begin{equation}\label{eq:KdVthomas1}
\left\{
\begin{array}{rcl}
\underline{u_t} & = & 0\,,\\[0.2em]
\underline{u_{x,x,x}} - 6 \, u \, u_x & = & 0\,,\\[0.2em]
u & \neq & 0\,,\\[0.2em]
u_{x,x} & \neq & 0\,,
\end{array}
\right.
\end{equation}
\begin{equation}\label{eq:KdVthomas2}
\left\{
\begin{array}{rcl}
\underline{u_t} - 6 \, u \, u_x & = & 0\,,\\[0.2em]
u_{x,x} & = & 0\,,\\[0.2em]
u & \neq & 0\,,
\end{array}
\right.
\end{equation}
\begin{equation}\label{eq:KdVthomas3}
\left\{
\begin{array}{rcl}
u & = & 0\,.
\end{array}
\right.
\end{equation}
In the above systems the highest ranked partial derivatives are underlined
(where not obvious).
We conclude that the set of power series solutions (\ref{eq:Taylor})
of (\ref{eq:KdV}) is the disjoint union of the sets of power series solutions
of the simple systems (\ref{eq:KdVthomas1}), (\ref{eq:KdVthomas2}), (\ref{eq:KdVthomas3}).
For (\ref{eq:KdVthomas1}) we can choose the values
of $a_{0,0}, \, a_{0,1}, \, a_{0,2} \in \C$ subject to the condition
that then $u(t,x)$ and its second derivative with respect to $x$ are
not identically zero. Then all other Taylor coefficients are uniquely determined
through the equations in (\ref{eq:KdVthomas1}):
\[
\begin{array}{rcll}
a_{j,k} \! & \! = \! & \! 0 & \qquad \mbox{for all } j \ge 1\,, \, k \ge 0\,,\\[0.5em]
a_{0,3+n} \! & \! = \! & \! \displaystyle
6 \, \sum_{k=0}^n \binom{n}{k} a_{0,k} \, a_{0,n-k+1} & \qquad \mbox{for all } n \ge 0\,.
\end{array}
\]
For (\ref{eq:KdVthomas2}) we can choose the values of
$a_{0,0}$, $a_{0,1} \in \C$ subject to the condition that then $u(t,x)$
is not identically zero.
Then all other Taylor coefficients are uniquely determined by
\[
\begin{array}{rcll}
a_{j,k} \! & \! = \! & \! 0 & \quad \mbox{for all } j \ge 0\,, \, k \ge 2\,,\\[1em]
a_{1+m,n} \! & \! = \! & \! \displaystyle
6 \, \sum_{j=0}^m \sum_{k=0}^n \binom{m}{j} \binom{n}{k} a_{j,k} \, a_{m-j,n-k+1} & \\[1.75em]
\! & \! = \! & \! \displaystyle
6^{1+m} \, (1+m)! \, a_{0,0}^{1-n} \, a_{0,1}^{1+m+n} & \quad \mbox{for all } m \ge 0\,, \, n \in \{ 0, 1 \}\,.
\end{array}
\]

\section{The theory of simple differential systems}
\label{sec:theory}

We recall the basic principles of simple differential systems and Thomas decomposition.
For more details, we refer to \cite{Robertz6}.

We consider partial differential equations of the form $p = 0$
as well as inequations of the form $q \neq 0$, where $p$ and $q$
are polynomials in unknown functions $u_1$, $u_2$, \ldots, $u_m$
of independent variables $x_1$, $x_2$, \ldots, $x_n$
and their partial derivatives. Since the differentiation order of
the equations to be dealt with by the Thomas decomposition method
is not precisely known before applying the method to the given PDE system,
the set of all expressions potentially occurring as left hand sides
is the smallest polynomial ring containing all partial derivatives
of $u_1$, $u_2$, \ldots, $u_m$, viz., the \emph{differential polynomial ring}
\[
R \, := \, \Q\{ u_1, \ldots, u_m \} \, := \, \Q[\, (u_k)_J \mid k \in \{ 1, \ldots, m \}, \, J \in (\Z_{\ge 0})^n \,]\,,
\]
where the \emph{differential indeterminate} $u_k = (u_k)_{(0, \ldots, 0)}$ represents the
unknown function $u_k(x_1, \ldots, x_n)$ with the same name and, more generally, $(u_k)_J$
for the multi-index $J = (j_1, j_2, \ldots, j_n) \in (\Z_{\ge 0})^n$ represents the partial derivative
\[
\frac{\partial^{j_1+j_2+\ldots+j_n} u_k}{\partial x_1^{j_1} \partial x_2^{j_2} \ldots \partial x_n^{j_n}}\,.
\]
The polynomial ring $R$ is closed under the derivations
$\partial_1$, \ldots, $\partial_n$ acting as
\[
\partial_i \, (u_k)_J \, := \, (u_k)_{J + 1_i}\,,
\qquad J + 1_i \, := \, (j_1, \ldots, j_{i-1}, j_i + 1, j_{i+1}, \ldots, j_n)\,,
\]
with additive extension to $R$ respecting the product rule of differentiation.
Hence, $\partial_i$ acts as the partial differential operator $\partial / \partial x_i$.
The coefficient field $\Q$ of rational numbers can also be replaced with a
larger field containing $\Q$ admitting $n$ derivations of which then the
derivations $\partial_i$ are extensions to $R$.
For example, the coefficient field can be chosen to be the field $\Q(x_1, \ldots, x_n)$
of rational functions in $x_1$, \ldots, $x_n$ with the usual derivations, if the
system to be dealt with consists of PDEs with rational function coefficients.

Given a system of partial differential equations
\begin{equation}\label{eq:diffsystem}
p_1 = 0\,, \quad
p_2 = 0\,, \quad
\ldots, \quad
p_s = 0\,,
\end{equation}
where $p_i \in R$, the consequences of (\ref{eq:diffsystem}) obtained by
taking linear combinations and derivatives
form a \emph{differential ideal} of $R$, viz., a (non-empty) subset of $R$
which is closed under taking linear combinations of its elements
with coefficients in $R$ and under differentiation (i.e.,
application of $\partial_1$, $\partial_2$, \ldots, $\partial_n$).
The differential ideal in question is said to be \emph{generated} by
the left hand sides $p_1$, $p_2$, \ldots, $p_s$.

Not every differential ideal of $R$ admits a finite set of generators.
However, the differential ideals corresponding to differential systems in a
Thomas decomposition are finitely generated
(cf.\ Theorems~\ref{thm:basisthm} and \ref{thm:vanish} below), which ensures termination of the decomposition algorithm (cf.~\cite{Robertz6}).
Differential ideals of this kind arise as follows.

Let $p$ be a differential polynomial in the differential indeterminate $u$
and $f$ be a complex analytic function of $x_1$, \ldots, $x_n$.
We write $p(f)$ for the result of substituting $f$ and
its partial derivatives $\partial^J f$
for $u$ and $u_J$ in $p$, respectively, $J \in (\Z_{\ge 0})^n$.
Similar notation can be introduced when substituting $m$ analytic functions
$f_1$, \ldots, $f_m$ for $m$ differential indeterminates $u_1$, \ldots, $u_m$.

Now, $(p^r)(f) = (p(f))^r = 0$ for some positive integer $r$ implies $p(f) = 0$.
A differential ideal of $R$ which contains for each of its elements $p$
also all differential polynomials in $R$ of which a power is equal to $p$
is said to be \emph{radical}. The differential ideal of all $p \in R$
which vanish under substitution of any analytic solution
of a PDE system is a radical differential ideal.
The following important theorem establishes a one-to-one correspondence
between radical differential ideals of $R$ and solutions sets (with complex analytic
functions on suitable domains) of PDE systems which are defined over $R$.

\begin{theorem}[Nullstellensatz of Ritt-Raudenbush, \cite{Ritt}, Sects.~II.7--11, IX.27]\label{thm:nullstellensatz}
Let $I$ be the differential ideal of $R$ generated by the left hand sides
$p_1$, $p_2$, \ldots, $p_s$ of a PDE system.
If a differential polynomial $p \in R$ vanishes under
substitution of any analytic solution of $p_1 = 0$, \ldots, $p_s = 0$,
then some power of $p$ is an element of $I$.
\end{theorem}

Radical differential ideals are finitely generated in the following sense.

\begin{theorem}[Basis Theorem of Ritt-Raudenbush, \cite{Ritt}, Sects.~I.12--14]\label{thm:basisthm}
%Every radical differential ideal of $R$ has a finite generating set.
For every radical differential ideal $I$ of $R$ there exists a finite
subset $B$ of $I$ such that $I$ is the smallest radical differential
ideal of $R$ which contains $B$.
\end{theorem}

Each differential system in a Thomas decomposition corresponds to a
radical differential ideal (cf.\ Theorem~\ref{thm:vanish} below).
The notions of simple differential system and Thomas decomposition
require a choice of a total ordering on the symbols representing the
unknown functions and their partial derivatives.

\begin{definition}\label{de:ranking}
A \emph{ranking} $\succ$ on $R$ is a total ordering on
\[
\Theta u \, := \, \{ \, (u_k)_J \mid k \in \{ 1, \ldots, m \}, \, J \in (\Z_{\ge 0})^n \, \}
\]
such that the following two conditions are satisfied, which express that
derivatives of an unknown function are ranked higher than the unknown function
itself and that $\succ$ is compatible with applying derivations.
\begin{enumerate}
\item For all $i \in \{ 1, \ldots, n \}$ and $k \in \{ 1, \ldots, m \}$ we have
$\partial_i \, u_k \succ u_k$.
\item For all
$i \in \{ 1, \ldots, n \}$, $k$, $l \in \{ 1, \ldots, m \}$, $K$, $L \in (\Z_{\ge 0})^n$,
the implication $(u_k)_K \succ (u_l)_L \, \Rightarrow \, \partial_i \, (u_k)_K \succ \partial_i \, (u_l)_L$ holds.
\end{enumerate}
\end{definition}

Examples of rankings were given in Section~\ref{sec:thomas}.
Of special interest to us are Riquier rankings.

\begin{definition}\label{def_riquier}
We define a {\em Riquier ranking} to be a ranking $\succ$ such that 
\begin{equation*}
\forall\, \delta_1,\delta_2\in \Theta:=\left\{ \,
\frac{\partial^{j_1+\cdots +j_n}}{\partial x_1^{j_1}\cdots \partial x_n^{j_n}} \; \middle| \; j_1,\ldots,j_n\in \mathbb{Z}_{\geq 0} \, \right\}
\end{equation*}
we have $\delta_1 u \succ \delta_2 u$ if the total differentiation order in $\delta_1 u$
is greater than the one in $\delta_2 u$ for any dependent variable $u$,
and the following condition holds for all dependent variables $v$, $w$:
\begin{equation*}\label{RiquierRanking}
\delta_1 v \succ \delta_2 v\quad \Longrightarrow\quad \delta_1 w \succ \delta_2 w\,.
\end{equation*}
\end{definition}

Assume that a ranking $\succ$ on $R$ has been chosen. Then every non-constant polynomial $p \in R$
involves a symbol $(u_k)_J$ which is maximal with respect to $\succ$.
It is referred to as the \emph{leader} $\ld(p)$ of $p$.
The differential polynomial $p$ can be represented recursively as a polynomial in $\ld(p)$
with coefficients which are polynomials in their leaders, etc. The coefficient of the highest
power of $\ld(p)$ in $p$ is called \emph{initial} of $p$ and denoted by $\initial(p)$.
The formal derivative of $p$ with respect to $\ld(p)$ is called the \emph{separant} of $p$
and denoted by $\separant(p)$.

A reduction process for differential polynomials can now be introduced as follows.
Let $p_1$ and $p_2$ be non-constant differential polynomials in $R$.
In case $\ld(p_1) = \ld(p_2)$ we consider the degrees $d_1$ and $d_2$ of $p_1$ and $p_2$
in $v := \ld(p_1)$, respectively. If $d_1 \ge d_2$, then
\[
\initial(p_2) \, p_1 - \initial(p_1) \, v^{d_1 - d_2} \, p_2
\]
is either constant, or has a leader which is ranked lower than $v$,
or has the same leader, but has smaller degree in $\ld(p_1)$ than $p_1$. If $d_1 < d_2$,
no reduction of $p_1$ modulo $p_2$ is possible.
If $\ld(p_1) \neq \ld(p_2)$, but there exists $J \in (\Z_{\ge 0})^n$
such that $v := \ld(p_1) = \partial^J \, \ld(p_2)$, then
\[
\separant(p_2) \, p_1 - \initial(p_1) \, v^{d_1 - 1} \, \partial^J \, p_2
\]
is either constant or has a leader which is ranked lower than $v$
(because $\partial^J \, p_2$ has degree one in $v$.)
If no $J \in (\Z_{\ge 0})^n$ exists such that $\ld(p_1) = \partial^J \, \ld(p_2)$, then
no reduction of $p_1$ modulo $p_2$ is possible.

This reduction process can be adapted so as to eliminate any occurrence
(in sufficiently high degree) of symbols $(u_k)_J$ which are leaders or
derivatives of leaders of $p_1$, \ldots, $p_s \in R$
in a given differential polynomial $p \in R$.
We say that $p \in R$ \emph{reduces to zero modulo $p_1$, \ldots, $p_s \in R$ and their derivatives}
if $p$ can be reduced to the zero polynomial in this way.

\begin{definition}\label{de:simplesystem}
A system of partial differential equations and inequations
\begin{equation}\label{eq:simplesys}
p_1 = 0\,, \quad \ldots\,, \quad p_s = 0\,, \quad q_1 \neq 0\,, \quad \ldots\,, \quad q_t \neq 0\,,
\quad \, \, \, (s, t \in \Z_{\ge 0})
\end{equation}
where $p_1$, \ldots, $p_s$, $q_1$, \ldots, $q_t$ are non-constant differential polynomials
in $R$, is said to be \emph{simple} (with respect to $\succ$)
if the following conditions are satisfied.
\begin{enumerate}
\item \label{de:simple:1} 
The leaders of $p_1$, $p_2$, \ldots, $p_s$, $q_1$, $q_2$, \ldots, $q_t$ are pairwise different.
\item \label{de:simple:2}
Let $v_1 \succ v_2 \succ \ldots \succ v_k$ be the elements of $\Theta u$ which
effectively occur in the differential polynomials $p_1$, \ldots, $p_s$, $q_1$, \ldots, $q_t$. We consider
(\ref{eq:simplesys}) as a system of polynomial equations and inequations
in $v_1$, $v_2$, \ldots, $v_k$.
If $r \in \{ p_1, \ldots, p_s, q_1, \ldots, q_t \}$ has leader $v_{\ell}$, thus
is a polynomial $r(v_{\ell}, v_{\ell+1}, \ldots, v_k)$, then we require that for every
solution $(a_1, a_2, \ldots, a_k) \in \C^k$ of (\ref{eq:simplesys})
the polynomial $r(v_{\ell}, a_{\ell+1}, a_{\ell+2}, \ldots, a_k)$
has the same degree in $v_{\ell}$ as $r(v_{\ell}, v_{\ell+1}, \ldots, v_k)$
and has no multiple roots.
(Equivalently, the initial of $r$ and the discriminant of $r$ with respect to $v_{\ell}$
do not vanish on the solution set of (\ref{eq:simplesys}) in $\C^k$.)
\item \label{de:simple:3}
The differential consequences of $p_1 = 0$, $p_2 = 0$, \ldots, $p_s = 0$ contain all
integrability conditions of this PDE system, i.e., the cross-derivative of each pair of distinct equations
whose leaders involve the same unknown function reduces to zero modulo $p_1$, \ldots, $p_s$
and their derivatives (passivity or formal integrability).
\item \label{de:simple:4}
No reduction of $q_1$, $q_2$, \ldots, $q_t$ is possible modulo $p_1$, $p_2$, \ldots, $p_s$ and
their derivatives.
\end{enumerate}
\end{definition}

\begin{definition}\label{de:thomasdecomposition}
Let $S$ be a system of partial differential equations and inequations, defined over $R$.
A \emph{Thomas decomposition} of $S$ (with respect to $\succ$) is a finite collection of
simple differential systems $S_1$, \ldots, $S_r$, defined over $R$, such that the solution sets
of $S_1$, \ldots, $S_r$ form a partition of the solution set of $S$, i.e., the solution set of $S$
is the disjoint union of the solution sets of $S_1$, \ldots, $S_r$.
\end{definition}

One should note that a Thomas decomposition of a differential system $S$ is not uniquely determined in general.

The method outlined above allows to compute a Thomas decomposition for any differential system $S$ as considered above, with respect to any ranking $\succ$, in finitely many steps \cite{BaechlerGerdtLangeHegermannRobertz2}, \cite{Robertz6}.
Case distinctions are necessary in general to ensure that initials and discriminants do not vanish on the solution set. Such a splitting into cases is performed by introducing the equation or inequation stating that the initial or discriminant vanishes or does not vanish, respectively. Hence, by construction, the simple differential systems in the resulting Thomas decomposition have disjoint solution sets.

The relevance of simple differential systems and the decomposition of a general
differential system into simple differential systems is explained by the following theorem.

\begin{theorem}[\cite{Robertz6}, Prop.~2.2.50]\label{thm:vanish}
Let a simple differential system $S$ be given by
\begin{equation}\label{Eq:simple_system}
p_1 = 0\,, \quad \ldots\,, \quad p_s = 0\,, \quad q_1 \neq 0\,, \quad \ldots\,, \quad q_t \neq 0\,,
\quad \, \, \, (s, t \in \Z_{\ge 0})
\end{equation}
where $p_1$, \ldots, $p_s$, $q_1$, \ldots, $q_t \in R$. Let $E$ be the differential ideal of $R$
which is generated by $p_1$, $p_2$, \ldots, $p_s$. Moreover, let $q \in R$ be the product of the
initials and separants of $p_1$, $p_2$, \ldots, $p_s$. Then the differential ideal
\[
E : q^{\infty} \, := \, \{ \, p \in R \mid q^r \, p \in E \mbox{ for some } r \in \Z_{\ge 0} \, \}
\]
is equal to the set of differential polynomials in $R$ which vanish under substitution of any
analytic solution of $S$. In particular, it is a radical differential ideal (and is therefore
finitely generated in the sense of Theorem~\ref{thm:basisthm}).
A differential polynomial $p$ is an element of $E : q^{\infty}$ if and only if $p$ reduces to
zero modulo $p_1$, \ldots, $p_s$ and their derivatives.
\end{theorem}

Note that while the solution sets of the simple differential systems $S_1$, \ldots, $S_r$ in
a Thomas decomposition are disjoint, the solution sets of the vanishing ideals
$E_1 : (q_1)^{\infty}$, \ldots, $E_r : (q_r)^{\infty}$ corresponding to $S_1$, \ldots, $S_r$
as considered in Theorem~\ref{thm:vanish}
are in general not disjoint (cf.\ also the Nullstellensatz, Theorem~\ref{thm:nullstellensatz}).

\begin{remark}\label{rem:parametricderivatives}
Let $S$ be a simple differential system as in Theorem~\ref{thm:vanish}.
For each $k \in \{ 1, \ldots, m \}$ let
$\partial^{J_{k,1}}$, $\partial^{J_{k,2}}$, \ldots, $\partial^{J_{k,n_k}}$,
where $J_{k,i} \in (\Z_{\ge 0})^n$, $n_k \in \Z_{\ge 0}$,
be the differential operators such that $\partial^{J_{k,i}} \, u_k = \ld(p_{j_i})$
for some $j_i \in \{ 1, \ldots, s \}$.
Due to the characterization of the vanishing ideal given by Theorem~\ref{thm:vanish},
the set of \emph{principal derivatives}
\[
P \, := \, \bigcup_{k=1}^m \, \bigcup_{i=1}^{n_k} \,
\{ \, \partial^J \, \partial^{J_{k,i}} \, u_k \mid J \in (\Z_{\ge 0})^n \, \}
\]
consists of those elements $v \in \Theta u$ for which there exists an equation with leader $v$ that is a consequence of $S$.
We refer to the elements of the complement $\Theta u \setminus P$ as the \emph{parametric derivatives}.
Note that $\ld(q_j) \in \Theta u \setminus P$ for all $j = 1$, \ldots, $t$
because of conditions~\ref{de:simple:1} and
\ref{de:simple:4} of Definition~\ref{de:simplesystem}.
\end{remark}

\bigskip

The differential Thomas decomposition is an appropriate tool to study and construct  formal and convergent power series solutions (cf.\ \cite[\S1.4]{LangeHegermannThesis}).

For a formal statement, we need two, generically satisfied, regularity assumptions.
Let $S$ be a simple differential system as in Theorem~\ref{thm:vanish} over a coefficient field of meromorphic functions in $n$ complex variables $x_1,\ldots,x_n$.
We call $x^0=(x^0_1, \ldots, x^0_n)\in\C^n$ a \emph{regular point} of $S$ if all coefficients of $p_i$ and $q_j$ are holomorphic and no $\initial(p_i), \separant(p_i), q_j$ vanishes at $x^0$ for all $1\le i\le s, 1\le j\le t$.
Consider formal power series $f=(f_1,\ldots,f_m)$ given by
\begin{align*}
f_k \, \, := \, \, \sum_{J \in (\Z_{\ge 0})^n} c_{k,J} \, \frac{(x_1 - x^0_1)^{J_1}}{J_1!} \ldots\frac{(x_n - x^0_n)^{J_n}}{J_n!}
\end{align*}
around a regular point $x^0$ of $S$.
We call any assignment $c_{k,J}\in\C$ where $\partial^J u_k$ are the parametric derivatives with $q_j(f_1, \ldots, f_m)(x^0_1,\ldots,x^0_n)\not=0$ for $j=1,\ldots,t$ a \emph{regular initial condition} of $S$ at the point $x^0$.

\begin{theorem}\label{thm:parametricderivatives}
  Let $S$ be a simple differential system and $x^0$ a regular point of $S$.
  Any regular initial condition of $S$ at $x^0$ completes to a formal power series solution of $S$ around $x_0$.
  For each such regular initial condition there are exactly as many completions as the product of the degrees of the equations of $S$ in their leaders.
  In particular, if all equations in $S$ are quasilinear, then there exists a unique such completion.
\end{theorem}

There may exist additional formal power series solutions that stem from non-regular initial conditions (cf.\ \cite[\S1.4]{LangeHegermannThesis}, \cite[Ex.~4.8, Ex.~4.9]{DiffCountPoly}).

By the results of Riquier \cite{Riquier}, this theorem holds for convergent power series solutions and, hence, generalizes Theorem~\ref{thm:cauchy-kovalevskaya} of Cauchy and Kovalevskaya (cf.\ also \cite{ThomasRiquier1}, \cite{ThomasRiquier2}, \cite{GerdtCauchyProblem}, \cite{RustReidWittkopf}).
We call a regular initial condition of $S$ at $x^0$ \emph{C-regular}, if all $f_k$ are (locally) convergent when setting all $c_{k,J}$ to zero for all principal derivatives $\partial^J u_k$.

\begin{cor}
  Let $S$ be a simple differential system w.r.t.\ a Riquier ranking and $x^0$ a regular point of $S$.
  If in Theorem~\ref{thm:parametricderivatives} the regular initial condition is C-regular, then any completion to a formal power series solution is also (locally) convergent.
\end{cor}

\bigskip

A ranking $\succ$ on $R = \Q\{ u_1, \ldots, u_m \}$ is called an \emph{elimination ranking}
with \emph{blocks} $B_1$, $B_2$, \ldots, $B_r$
if $B_1 \cup B_2 \cup \ldots \cup B_r$ is a partition of the set $\{ u_1, \ldots, u_m \}$
and we have
\[
\partial^{K} u_{i} \succ \partial^{L} u_{j}\,,
\qquad K, L \in (\Z_{\ge 0})^n\,,
\]
whenever $u_{i} \in B_{k}$ and $u_{j} \in B_{l}$ with $k < l$.

\medskip

By computing a Thomas decomposition with respect to an elimination ranking one may
find all consequences of a differential system involving only differential indeterminates
which are in lower ranked blocks.
More precisely, we have the following theorem.

\begin{theorem}[\cite{Robertz6}, Prop.~3.1.36]\label{thm:diffelim}
Let $S$ be a simple differential system with respect to an elimination ranking $\succ$
with blocks $B_1$, $B_2$, \ldots, $B_r$.
Let $E$ and $q$ be defined as in Theorem~\ref{thm:vanish},
and for all $i \in \{ 1, \ldots, r \}$ let $E_i$ be the differential ideal
of $\Q\{ \, u_j \mid u_j \in B_1 \cup B_2 \cup \ldots \cup B_i \, \}$
which is generated by the left hand sides of the equations in $S$ which only involve
differential indeterminates in $B_1 \cup B_2 \cup \ldots \cup B_i$,
and let $q_i$ be the product of the initials and separants of these differential polynomials.
Then for every $i \in \{ 1, \ldots, r \}$ we have
\[
(E : q^{\infty}) \cap \Q\{ \, u_j \mid u_j \in B_1 \cup B_2 \cup \ldots \cup B_i \, \} =
E_i : q_i^{\infty}\,.
\]
\end{theorem}

\section{Package commands}
\label{sec:commands}

The ranking, the total order on the set of differential indeterminates, is globally determined by the command \texttt{ComputeRanking(ivar,dvar)}.
Here, \texttt{ivar} is list of (lists of) independent variables, and \texttt{dvar} is a list of (lists of) unknown functions.
The unknown functions can be grouped in lists to use a block elimination ranking, e.g.\ \texttt{dvar=[[u1],[u2,u3]]} represents \texttt{u1} and all its derivatives being bigger than \texttt{u2} or \texttt{u3} or any of their derivatives.
In case two unknown functions are equally big, we can order their derivatives by grouping the independent variables in lists, e.g.\ \texttt{ivar=[[x],[y,z]]} means that a differential indeterminate ranks higher if its order w.r.t.\ $x$ is bigger; in case of a tie the oder in $y$ and $z$ is considered.
If no additional grouping is done, the degree-reverse lexicographical ranking (cf.\ \eqref{eq_ranking}) is chosen.

For the remainder of this section, we assume the following ranking.
\medskip

\begin{mapleinput}
\mapleinline{active}{1d}{ComputeRanking([x,y], [u,v]);}{}
\end{mapleinput}
\medskip

\noindent
Derivatives of the unknown functions are encoded by indexing them  with the order vector as index, e.g.\ $\mathtt{u[2,1]}\equiv u_{x,x,y}\equiv \frac{\partial^3}{\partial x^2 \partial y}u(x,y)$.
The coefficients can be any valid Maple expressions.
In particular, functions in the independent variables are correctly differentiated.
Conversion from and to the \texttt{diff} notation of Maple works via \texttt{Diff2JetList} and \texttt{JetList2Diff}.
\medskip

\begin{mapleinput}
\mapleinline{active}{1d}{p := Diff2JetList(u(x,y)+v(x,y)*diff(u(x,y),x,x,y));}{}
\end{mapleinput}
\begin{maplelatex}
\mapleinline{inert}{2d}{}{$p:=v_{0, 0} u_{2, 1} + u_{0, 0}$}
\end{maplelatex}
\begin{mapleinput}
\mapleinline{active}{1d}{JetList2Diff(p);}{}
\end{mapleinput}
\begin{maplelatex}
\mapleinline{inert}{2d}{}{$u\!\left(x,y\right) + v\!\left(x,y\right) \left({\frac {\partial ^{3}}{\partial {x}^{2}\partial {y}}}u\!\left(x,y\right)\right)$}
\end{maplelatex}
\medskip

\noindent
We can compute (partial) derivatives, initial, leader, and separant of a differential polynomial.
\medskip

\begin{mapleinput}
\mapleinline{active}{1d}{PartialDerivative(p, x, y);
}{}
\end{mapleinput}
\begin{maplelatex}
\mapleinline{inert}{2d}{}{$ u_{2, 1}v_{1, 1}+u_{2, 2}v_{1, 0}+u_{3, 1}v_{0, 1}+u_{3, 2}v_{0, 0}+u_{1, 1}$}
\end{maplelatex}

\begin{mapleinput}
\mapleinline{active}{1d}{Leader(p);
}{}
\end{mapleinput}
\begin{maplelatex}
\mapleinline{inert}{2d}{}{$u_{2, 1}$}
\end{maplelatex}

\begin{mapleinput}
\mapleinline{active}{1d}{Initial(p);
}{}
\end{mapleinput}
\begin{maplelatex}
\mapleinline{inert}{2d}{}{$v_{0, 0}$}
\end{maplelatex}

\begin{mapleinput}
\mapleinline{active}{1d}{Separant(p);
}{}
\end{mapleinput}
\begin{maplelatex}
\mapleinline{inert}{2d}{}{$v_{0, 0}$}
\end{maplelatex}
\medskip

\noindent
The command \texttt{DifferentialThomasDecomposition} computes a Thomas decomposition.
Its first parameter is a list of (left hand sides of) differential equations and its second parameter is a list of (left hand sides of) differential inequations.
It returns a Thomas decomposition as a list of differential systems.
\medskip

\begin{mapleinput}
\mapleinline{active}{1d}{res := DifferentialThomasDecomposition([p], []);
}{}
\end{mapleinput}
\begin{maplelatex}
\mapleinline{inert}{2d}{}{$ {\it res}\, := \,[{\it DifferentialSystem},{\it DifferentialSystem}]$}
\end{maplelatex}
\medskip

\noindent
The commands \texttt{DifferentialSystemEquations} resp.\ \texttt{DifferentialSystemInequations} resp. \texttt{PrettyPrintDifferentialSystem} extract equations resp.\ inequations resp.\ both from a given differential system.
\medskip

\begin{mapleinput}
\mapleinline{active}{1d}{PrettyPrintDifferentialSystem(res[1]);}{}
\end{mapleinput}
\begin{maplelatex}
\mapleinline{inert}{2d}{}{$[ \,
\left( {\frac {\partial ^{3}}{\partial x^2 \, \partial y}} \, u(x,y) \right) v(x,y) + u(x,y) = 0, \,
v(x,y) \neq 0 \, ]$}
\end{maplelatex}
\medskip

\noindent
\begin{mapleinput}
\mapleinline{active}{1d}{PrettyPrintDifferentialSystem(res[2]);}{}
\end{mapleinput}
\begin{maplelatex}
\mapleinline{inert}{2d}{}{$[ \,
u(x,y)=0 , \,
v(x,y)= 0 \, ]$}
\end{maplelatex}
\medskip

\noindent
A differential system can be passed to the differential equation solvers of Maple via \texttt{MyPDSolve}.
\medskip

\begin{mapleinput}
\mapleinline{active}{1d}{map(MyPDSolve, res);}{}\noindent
\end{mapleinput}
\begin{maplelatex}
\mapleinline{inert}{2d}{}{$
\begin{array}{c}
[ \,
  \left\{ u(x,y) = 0, \, v(x,y) = v(x,y) \right\},  
  \left\{ u(x,y) = u(x,y), \, v(x,y) = -{\frac {u(x,y) }{{\frac {\partial ^{3}}{\partial y\partial {x}^2}} u(x,y) }} \right\},\\[1em]
  \left\{ u(x,y) =0, \, v(x,y) = 0 \right\}
\, ]
\end{array}$}
\end{maplelatex}
\medskip

\noindent
The command \texttt{IntersectDecompositions} computes the intersection of two lists of disjoint systems.
\medskip

\begin{mapleinput}
\mapleinline{active}{1d}{IntersectDecompositions(res,
\quad DifferentialThomasDecomposition([u[0,0]-v[0,0]], []));}{}
\end{mapleinput}\noindent
\begin{maplelatex}
\mapleinline{inert}{2d}{}{$ [{\it DifferentialSystem},{\it DifferentialSystem}]$}
\end{maplelatex}
\medskip

\begin{mapleinput}
\mapleinline{active}{1d}{PrettyPrintDifferentialSystem(%);}{}
\end{mapleinput}
\begin{maplelatex}
\mapleinline{inert}{2d}{}{$
\begin{array}{c}
[ \, [ \, u(x,y) -v(x,y) = 0, \,
{\frac {\partial ^{3}} {\partial y\partial {x}^{2}}} \, v(x,y) +1 = 0, \, v(x,y) \neq 0 \, ],\\[0.5em]
[ \, u(x,y) = 0, \, v \left( x,y \right) =0 \, ] \, ]
\end{array}$}
\end{maplelatex}
\medskip

Both \texttt{DifferentialSystemReduce} and \texttt{DifferentialSystemNormalForm} bring (list of) differential equation(s) in a reduced form w.r.t.\ (list of) differential system(s).
This reduced form is zero iff the differential equation is a consequence of the system.
The output is a (list of (lists of)) differential equation(s), as each differential equation is treated with each differential system.
\texttt{DifferentialSystemReduce} applies a differential reduction, hence the resulting differential equation is equivalent to the input differential equation up to a factor, which is implied to be non-zero by the system.
\texttt{DifferentialSystemNormalForm} additionally divides by this factor to have an equivalent differential equation, at the cost of denominators.
\medskip

\begin{mapleinput}
\mapleinline{active}{1d}{DifferentialSystemReduce(
\quad [PartialDerivative(p,x),PartialDerivative(p,y),u[1,1]],
\quad res);}{}
\end{mapleinput}
\noindent
\begin{maplelatex}
\mapleinline{inert}{2d}{}{$[ \, [0,0],[0,0],[u_{{1,1}},0] \, ]$}
\end{maplelatex}
\medskip

\noindent
The dimension polynomial \texttt{DifferentialSystemDimensionPolynomial} is a univariate polynomial function \cite{DiffDimPoly}.
When evaluated on any high enough order, it yields the number of generically freely choosable power series coefficients of a solution of a simple differential system up to that order.
It is a numerical polynomial and, hence, can uniquely be represented as a linear combination of the binomial coefficients  $C_{s+i}^{\,i}$ for $0\le i\le n$ with positive interger coefficients by \texttt{DifferentialSystemDimensionPolynomialCanonicalBase}. 
\medskip

\begin{mapleinput}
\mapleinline{active}{1d}{DifferentialSystemDimensionPolynomial(res[1]);}{}
\end{mapleinput}
\begin{maplelatex}
\mapleinline{inert}{2d}{}{$\frac{1}{2}{s}^{2}+\frac{9}{2}s+1$}
\end{maplelatex}
\medskip

\begin{mapleinput}
\mapleinline{active}{1d}{DifferentialSystemDimensionPolynomialCanonicalBase(res[1]);}{}
\end{mapleinput}
\begin{maplelatex}
\mapleinline{inert}{2d}{}{$\binom{2+s}{s}+3s$}
\end{maplelatex}
\medskip

\noindent
The command \texttt{PowerSeriesSolution} allows to compute the set of generic power series solutions up to a given order around a given point.
\medskip

\begin{mapleinput}
\mapleinline{active}{1d}{PowerSeriesSolution(res[1], 2, [a,b]);}{}
\end{mapleinput}
\begin{maplelatex}
\mapleinline{inert}{2d}{}{$[u_{{0,0}}+u_{{1,0}} \left( x-a \right) +u_{{0,1}} \left( y-b \right) ,v_{{0,0}}+v_{{1,0}} \left( x-a \right) +v_{{0,1}} \left( y-b \right) ]$}
\end{maplelatex}

%DifferentialSystemEquationsAndReductiveProlongations(TD[1]);

\section{Examples}
\label{sec:examples}

\begin{example}[Consistency check]
This example illustrates the verification of consistency for differential systems.
We consider the following systems of PDEs for one unknown function $u(x,y)$ of the two independent variables $x,y$:
\[
u_xu_y+u_x+1=0\,,\quad uu_{x,x}-u_x-u_y^2+au=0\,,\quad a\in \R\,.
\]
We would like to detect all values of the parameter '$a$' for which the PDE system is consistent. In order to determine these values, we treat parameter '$a$' as a function of $x,y$ whose partial derivatives are identically zero.

\begin{maplegroup}
\begin{mapleinput}
\mapleinline{active}{1d}{with(DifferentialThomas):
}{}
\end{mapleinput}
\end{maplegroup}
\begin{maplegroup}
\begin{mapleinput}
\mapleinline{active}{1d}{ComputeRanking([x,y], [[u],[a]]):
}{}
\end{mapleinput}
\end{maplegroup}
\begin{maplegroup}
\begin{mapleinput}
\mapleinline{active}{1d}{DS := [u[1,0]*u[0,1]+u[1,0]+1,u[0,0]*u[2,0]-u[1,0]-u[0,1]^2+
}{}
\vskip 0.1cm
\mapleinline{active}{1d}{a[0,0]*u[0,0],a[1,0],a[0,1]]:
}{}
\end{mapleinput}
\end{maplegroup}
\begin{maplegroup}
\begin{mapleinput}
\mapleinline{active}{1d}{TD := DifferentialThomasDecomposition([DS, []);
}{}
\end{mapleinput}
\mapleresult
\begin{maplelatex}
\mapleinline{inert}{2d}{TD := [DifferentialSystem]}{\[\displaystyle {\it TD}\, := \,[{\it DifferentialSystem}]\]}
\end{maplelatex}
\end{maplegroup}
\begin{maplegroup}
\begin{mapleinput}
\mapleinline{active}{1d}{PrettyPrintDifferentialSystem(TD[1]);
}{}
\end{mapleinput}
\mapleresult
\begin{maplelatex}
\[
\begin{array}{c}
\displaystyle [ \,\left( {\frac {\partial }{\partial x}} \, u(x,y)\right)^3 + \left( {\frac {\partial }{\partial x}} \, u(x,y)\right)^2 +2\left( {\frac {\partial }{\partial x}} \, u(x,y)\right)+1 =0,\\[1em]
\displaystyle
\left( {\frac {\partial }{\partial y}} \, u(x,y)\right) \left( {\frac {\partial }{\partial x}} \, u(x,y)\right) + \left( {\frac {\partial }{\partial x}} \, u(x,y)\right)+1 =0,\,
a(x,y)=0,\, u(x,y) \neq 0 \, ]
\end{array}
\]
\end{maplelatex}
\end{maplegroup}
\smallskip

\noindent
Thus, the differential system under consideration is consistent if and only if $a=0$. Since every simple system has a solution~\cite{Thomas,BaechlerGerdtLangeHegermannRobertz2,Robertz6}, a  PDE system is consistent if and only if its Thomas decomposition is nonempty.  
\end{example}

%Arbitraryness and implied equations: Navier Stokes from the worksheet, make sure it includes the dimension polynomial and reduction of the pressure equations

\begin{example}
As a second example we consider the
Navier-Stokes equations for an incompressible flow of a constant viscosity fluid:
\begin{equation}\label{eq:navierstokes}
\left\{
\begin{array}{rcl}
\mathbf{u}_t + (\mathbf{u} \cdot \nabla) \, \mathbf{u} + \nabla p - \mu \, \Delta \mathbf{u} & = & 0\,,\\[0.2em]
\nabla \cdot \mathbf{u} & = & 0\,,
\end{array}
\right.
\end{equation}
where $\mathbf{u}=(u,v,w)$ is the velocity vector,
$p$ the pressure, and $\mu = 1 / {\rm Re}$ for the Reynolds number ${\rm Re}$.
In Cartesian coordinates $x$, $y$, $z$ of $\R^3$ we have, equivalently,
\begin{equation}\label{eq:navstcartesian}
\left\{
\begin{array}{rcl}
u_t + u \, u_x + v \, u_y + w \, u_z + p_x - \mu \, (u_{x,x} + u_{y,y} + u_{z,z}) & = & 0\,,\\[0.5em]
v_t + u \, v_x + v \, v_y + w \, v_z + p_y - \mu \, (v_{x,x} + v_{y,y} + v_{z,z}) & = & 0\,,\\[0.2em]
w_t + u \, w_x + v \, w_y + w \, w_z + p_z - \mu \, (w_{x,x} + w_{y,y} + w_{z,z}) & = & 0\,,\\[0.2em]
u_x + v_y + w_z & = & 0\,.
\end{array}
\right.
\end{equation}
We would like to verify algorithmically the well known fact that the Poisson pressure equation
\begin{equation}\label{eq:poisson}
\Delta p + \nabla \cdot ((\mathbf{u} \cdot \nabla) \, \mathbf{u}) \, \, = \, \, 0
\end{equation}
is a consequence of the Navier-Stokes equations (\ref{eq:navierstokes}).

We choose the degree-reverse lexicographical ranking $\succ$ on the set of
partial derivatives
\[
\left\{ \,
\frac{\partial^{i+j+k+l} f}{\partial t^i \, \partial x^j \, \partial y^k \, \partial z^l} \; \middle| \;
f \in \{ u, v, w, p \}, \;
i, j, k, l \in \Z_{\ge 0} \, \right\}
\]
which first compares the differential operators and in case of equality compares
the unknown functions to which the differential operators are applied according to $u \succ v \succ w \succ p$.
For system (\ref{eq:navstcartesian}) the resulting Thomas decomposition (with respect to $\succ$) consists
of only one simple differential system:
\[
\left\{
\begin{array}{rcl}
\underline{u_x}+v_y+w_z \! & \! = \! & \! 0\,,\\[0.2em]
\mu\,\underline{v_{x,x}}+\mu\,v_{y,y}+\mu\,v_{z,z}-uv_{x}-vv_{y}-wv_{z}-p_{y}-v_{t} \! & \! = \! & \! 0\,,\\[0.2em]
\mu\,u_{y,y}+\mu\,u_{z,z}-\mu\,\underline{v_{x,y}}-\mu\,w_{x,z}+uv_{y}+uw_{z}-vu_{y}-wu_{z}-p_{x}-u_{t} \! & \! = \! & \! 0\,,\\[0.2em]
\mu\,\underline{w_{x,x}}+\mu\,w_{y,y}+\mu\,w_{z,z}-uw_{x}-vw_{y}-ww_{z}-p_{z}-w_{t} \! & \! = \! & \! 0\,,\\[0.2em]
\underline{p_{x,x}}+p_{y,y}+p_{z,z}+2\,u_{y}v_{x}+2\,u_{z}w_{x}+2\,v_{y}^{2}+2\,v_{y}w_{z}+2\,v_{z}w_{y}+2\,w_{z}^{2} \! & \! = \! & \! 0\,,
\end{array}
\right.
\]
where some reductions have been performed on the first equation in (\ref{eq:navstcartesian})
to yield the second equation in the obtained simple system. The last equation was obtained as
\[
\frac{\partial}{\partial x} A_1 +
\frac{\partial}{\partial y} A_2 +
\frac{\partial}{\partial z} A_3 +
\left[ v \, \Delta - \frac{\partial}{\partial t} - u \frac{\partial}{\partial x} - v \frac{\partial}{\partial y}
-w \frac{\partial}{\partial z} + 2 \, v_y + 2 \, w_z \right] A_4\,,
\]
where $A_1$, $A_2$, $A_3$, $A_4$ are the left hand sides of the equations in (\ref{eq:navstcartesian})
and $\Delta$ is the Laplace operator.
%\[
%\Delta p + 2 \, u_y \, v_x + 2 \, u_z \, w_x + 2 \, v_y^2 + 2 \, v_y \, w_z + 2 \, v_z \, w_y + 2 \, w_z^2\,.
%\]
In fact, modulo the other equations in the system, the last equation 
is equivalent to the Poisson pressure equation (\ref{eq:poisson}).

The above simple system of the Thomas decomposition allows to enumerate the
Taylor coefficients of $u(t, x, y, z)$, $v(t, x, y, z)$, $w(t, x, y, z)$,
$p(t, x, y, z)$ whose values can be chosen arbitrarily to obtain
a power series solution of (\ref{eq:navierstokes}).
These correspond to the partial derivatives of $u$, $v$, $w$, $p$ which
are not highest ranked derivatives in any differential equation among the
consequences of the simple system.
For $u(t, x, y, z)$ this enumeration is obtained from the non-zero terms of the
formal power series expansion of
\[
\frac{1}{(1 - \partial_t) \, (1 - \partial_y) \, (1 - \partial_z)}\,,
\]
where $\partial_t = \partial / \partial t$,
%$\partial_x = \partial / \partial x$,
$\partial_y = \partial / \partial y$,
$\partial_z = \partial / \partial z$.
In other words, an arbitrary analytic function of $t$, $y$, $z$ can be
prescribed in terms of boundary conditions to determine $u(t, x, y, z)$.
For $v(t, x, y, z)$ the enumeration is given by
\[
\frac{1}{(1 - \partial_t) \, (1 - \partial_y) \, (1 - \partial_z)} +
\frac{\partial_x}{(1 - \partial_t) \, (1 - \partial_z)}
\]
and for $w(t, x, y, z)$ and for $p(t, x, y, z)$ by
\[
\frac{1}{(1 - \partial_t) \, (1 - \partial_y) \, (1 - \partial_z)} +
\frac{\partial_x}{(1 - \partial_t) \, (1 - \partial_y) \, (1 - \partial_z)}\,.
\]
Hence, extending the Cauchy-Kovaleskaya Theorem~\ref{thm:cauchy-kovalevskaya}
we may pose the Cauchy problem
for the Navier-Stokes equations (\ref{eq:navierstokes}) around an
arbitrary point $(t_0, x_0, y_0, z_0)$ as follows:
\[
\left\{ \begin{array}{rcl}
u(t, x_0, y, z) & = & f_1(t, y, z)\,,\\[0.5em]
v(t, x_0, y, z) & = & f_2(t, y, z)\,,\\[0.5em]
\displaystyle
\frac{\partial v}{\partial x}(t, x_0, y_0, z) & = & f_3(t, z)\,,\\[1em]
w(t, x_0, y, z) & = & f_4(t, y, z)\,,\\[0.5em]
\displaystyle
\frac{\partial w}{\partial x}(t, x_0, y, z) & = & f_5(t, y, z)\,,\\[1em]
p(t, x_0, y, z) & = & f_6(t, y, z)\,,\\[0.5em]
\displaystyle
\frac{\partial p}{\partial x}(t, x_0, y, z) & = & f_7(t, y, z)\,,
\end{array} \right.
\]
where $f_1$, $f_2$, \ldots, $f_7$ are arbitrary functions of their arguments which are analytic around the point $(t_0, x_0, y_0, z_0)$. The arbitrariness of analytic solutions to (\ref{eq:navierstokes}) is determined by $f_1$, $f_2$, \ldots, $f_7$.
\end{example}

\begin{example}
We demonstrate how to study the classical Cole-Hopf transformation by using the
differential Thomas decomposition (cf.\ also \cite[Ex.~3.8]{BaechlerGerdtLangeHegermannRobertz2}).
The claim is that for every non-zero analytic solution $\eta(t, x)$ of the heat equation
\begin{equation}\label{eq:heatequation}
u_t + u_{x,x} \, = \, 0
\end{equation}
the function
\begin{equation}\label{eq:ColeHopf}
\zeta(t, x) \, := \, \frac{\eta_x(t, x)}{\eta(t, x)}
\end{equation}
is a solution of Burgers' equation
\begin{equation}\label{eq:BurgersColeHopf}
u_t + u_{x,x} + 2 \, u \, u_x \, = \, 0\,.
\end{equation}
We define a ranking $\succ$ on the ring of differential polynomials in $\eta$ and $\zeta$
such that any partial derivative of $\eta$ is ranked higher with respect to $\succ$ than
any partial derivative of $\zeta$.
According to Theorem~\ref{thm:diffelim} this allows determining all differential equations
which are satisfied by $\zeta$ from a Thomas decomposition with respect to $\succ$.
\medskip

\begin{maplegroup}
\begin{mapleinput}
\mapleinline{active}{1d}{with(DifferentialThomas):
}{}
\end{mapleinput}
\end{maplegroup}
\begin{maplegroup}
\begin{mapleinput}
\mapleinline{active}{1d}{ComputeRanking([t,x], [[eta],[zeta]]):
}{}
\end{mapleinput}
\end{maplegroup}
\medskip

\noindent
We define the differential system which combines (\ref{eq:heatequation})
and (\ref{eq:ColeHopf}).
\medskip

\begin{maplegroup}
\begin{mapleinput}
\mapleinline{active}{1d}{CH := [eta[1,0]+eta[0,2], eta[0,0]*zeta[0,0]-eta[0,1]];
}{}
\end{mapleinput}
\mapleresult
\begin{maplelatex}
\mapleinline{inert}{2d}{CH := [eta[1, 0]+eta[0, 2], Zeta[0, 0]*eta[0, 0]-eta[0, 1]]}{\[\displaystyle {\it CH}\, := \, [ \, \eta_{{1,0}}+\eta_{{0,2}}, \, \eta_{{0,0}} \, \zeta_{{0,0}}-\eta_{{0,1}} \, ]
\]}
\end{maplelatex}
\end{maplegroup}
\medskip

\noindent
We also include the assumption $\eta \neq 0$ as an inequation.
\medskip

\begin{maplegroup}
\begin{mapleinput}
\mapleinline{active}{1d}{TD := DifferentialThomasDecomposition(CH, [eta[0,0]]);
}{}
\end{mapleinput}
\mapleresult
\begin{maplelatex}
\mapleinline{inert}{2d}{TD := [DifferentialSystem]}{\[\displaystyle {\it TD}\, := \,[{\it DifferentialSystem}]\]}
\end{maplelatex}
\end{maplegroup}
\begin{maplegroup}
\begin{mapleinput}
\mapleinline{active}{1d}{PrettyPrintDifferentialSystem(TD[1]);
}{}
\end{mapleinput}
\mapleresult
\begin{maplelatex}
\mapleinline{inert}{2d}{[eta(t, x)*Zeta(t, x)^2+eta(t, x)*(diff(Zeta(t, x), x))+diff(eta(t, x), t) = 0, eta(t, x)*Zeta(t, x)-(diff(eta(t, x), x)) = 0, 2*Zeta(t, x)*(diff(Zeta(t, x), x))+diff(Zeta(t, x), x, x)+diff(Zeta(t, x), t) = 0, eta(t, x) <> 0]}{
\[
\begin{array}{c}
\displaystyle [ \, {\frac {\partial }{\partial t}} \, \eta(t,x) + \left( \zeta(t,x) \right)^2 \eta(t,x) + \left( {\frac {\partial }{\partial x}} \, \zeta(t,x) \right) \eta(t,x) = 0, \quad
{\frac {\partial }{\partial x}} \, \eta(t,x) - \zeta(t,x) \, \eta(t,x) =0,\\[1em]
\displaystyle
{\frac {\partial^2}{\partial {x}^2}} \, \zeta(t,x) + {\frac {\partial }{\partial t}} \, \zeta(t,x) +
2\,\zeta(t,x) \, {\frac {\partial }{\partial x}} \, \zeta(t,x) =0, \, \eta(t,x) \neq 0 \, ]
\end{array}
\]}
\end{maplelatex}
\end{maplegroup}
\medskip

\noindent
The simple system of the resulting Thomas decomposition allows to read off that
$\zeta$ as defined by (\ref{eq:ColeHopf}) is a solution of (\ref{eq:BurgersColeHopf})
if $\eta$ is a solution of (\ref{eq:heatequation}), which proves the original claim.
Conversely, since the above simple differential system is consistent with the heat equation
(\ref{eq:heatequation}) for $\eta$ by construction, we conclude that for any solution
$\zeta$ of (\ref{eq:BurgersColeHopf}) there exists a solution $\eta$ of (\ref{eq:heatequation})
such that the Cole-Hopf transformation of $\eta$ is $\zeta$.
\end{example}

\begin{example}[Detection of hidden constraints for the system with singular Lagrangian]
We apply the differential Thomas decomposition to compute the hidden Lagrangian constraints for the problem taken from~\cite[Ex.~6.3]{GerdtRobertz5}, \cite[Eq.~(8.1)]{Deriglazov}.
The model Lagrangian reads
\[
L \, = \,
q_2^2 \, (q_1)_t^2 + q_1^2 \, (q_2)_t^2 + 2 \, q_1 \, q_2 \, (q_1)_t \, (q_2)_t + q_1^2 + q_2^2.
\]
We define $L$ and the corresponding Euler-Lagrange equations in Maple:
\medskip

\begin{maplegroup}
\begin{mapleinput}
\mapleinline{active}{1d}{L := q2[0]\symbol{94}2*q1[1]\symbol{94}2+q1[0]\symbol{94}2*q2[1]\symbol{94}2+2*q1[0]*q2[0]*q1[1]*q2[1]}{}
\vskip 0.1cm
\mapleinline{active}{1d}{+q1[0]\symbol{94}2+q2[0]\symbol{94}2; 
\vskip 0.2cm
EL := map(a->PartialDerivative(diff(L, a[1]), t)-diff(L, a[0]), dvar);}{}
\end{mapleinput}
\mapleresult
\begin{maplelatex}
\mapleinline{active}{2d}{L := VectorCalculus:-`+`(VectorCalculus:-`+`(VectorCalculus:-`+`(VectorCalculus:-`+`(VectorCalculus:-`*`(q1[0]^2, q2[1]^2), VectorCalculus:-`*`(VectorCalculus:-`*`(VectorCalculus:-`*`(VectorCalculus:-`*`(2, q1[0]), q2[0]), q1[1]), q2[1])), VectorCalculus:-`*`(q2[0]^2, q1[1]^2)), q1[0]^2), q2[0]^2)}{\[\displaystyle L\, := \,{{\it q1}_{{0}}}^{2}{{\it q2}_{{1}}}^{2}+2\,{\it q1}_{{0}}{\it q1}_{{1
}}{\it q2}_{{0}}{\it q2}_{{1}}+{{\it q1}_{{1}}}^{2}{{\it q2}_{{0}}}^{2}+{{\it q1}_{{0}}}^{2}+{{\it q2}_{{0}}}^{2}
 \]}
\end{maplelatex}
\mapleresult
\begin{maplelatex}
\mapleinline{active}{2d}{EL := [VectorCalculus:-`+`(VectorCalculus:-`+`(VectorCalculus:-`+`(VectorCalculus:-`*`(VectorCalculus:-`*`(VectorCalculus:-`*`(2, q1[0]), q2[0]), q2[2]), VectorCalculus:-`*`(VectorCalculus:-`*`(VectorCalculus:-`*`(4, q1[1]), q2[0]), q2[1])), VectorCalculus:-`*`(VectorCalculus:-`*`(2, q1[2]), q2[0]^2)), VectorCalculus:-`-`(VectorCalculus:-`*`(2, q1[0]))), VectorCalculus:-`+`(VectorCalculus:-`+`(VectorCalculus:-`+`(VectorCalculus:-`*`(VectorCalculus:-`*`(2, q1[0]^2), q2[2]), VectorCalculus:-`*`(VectorCalculus:-`*`(VectorCalculus:-`*`(4, q1[1]), q1[0]), q2[1])), VectorCalculus:-`*`(VectorCalculus:-`*`(VectorCalculus:-`*`(2, q1[0]), q2[0]), q1[2])), VectorCalculus:-`-`(VectorCalculus:-`*`(2, q2[0])))]}{
\[
\displaystyle EL\, := \,[ \, 2\,{\it q1}_{{0}}{\it q2}_{{0}}{\it q2}_{{2}}+4\,{\it q1}_{{1}}{\it 
q2}_{{0}}{\it q2}_{{1}}+2\,{\it q1}_{{2}}{{\it q2}_{{0}}}^{2}-2\,{\it q1}_{{0}},\,
\]}
\end{maplelatex}
\begin{maplelatex}
\mapleinline{active}{2d}{LE := [VectorCalculus:-`+`(VectorCalculus:-`+`(VectorCalculus:-`+`(VectorCalculus:-`*`(VectorCalculus:-`*`(VectorCalculus:-`*`(2, q1[0]), q2[0]), q2[2]), VectorCalculus:-`*`(VectorCalculus:-`*`(VectorCalculus:-`*`(4, q1[1]), q2[0]), q2[1])), VectorCalculus:-`*`(VectorCalculus:-`*`(2, q1[2]), q2[0]^2)), VectorCalculus:-`-`(VectorCalculus:-`*`(2, q1[0]))), VectorCalculus:-`+`(VectorCalculus:-`+`(VectorCalculus:-`+`(VectorCalculus:-`*`(VectorCalculus:-`*`(2, q1[0]^2), q2[2]), VectorCalculus:-`*`(VectorCalculus:-`*`(VectorCalculus:-`*`(4, q1[1]), q1[0]), q2[1])), VectorCalculus:-`*`(VectorCalculus:-`*`(VectorCalculus:-`*`(2, q1[0]), q2[0]), q1[2])), VectorCalculus:-`-`(VectorCalculus:-`*`(2, q2[0])))]}{
\[
\displaystyle 2\,{{\it q1}_{{0}}}^{2}{\it q2}_{{2}}+4\,{\it q1}_{{0}}{\it 
q1}_{{1}}{\it q2}_{{1}}+2\,{\it q1}_{{0}}{\it q1}_{{2}}{\it q2}_{{0}}-
2\,{\it q2}_{{0}} \, ]
\]}
\end{maplelatex}
\end{maplegroup}
\begin{maplegroup}
\begin{mapleinput}
\mapleinline{active}{1d}{with(DifferentialThomas):
}{}
\end{mapleinput}
\end{maplegroup}
\begin{maplegroup}
\begin{mapleinput}
\mapleinline{active}{1d}{ivar := [t]: dvar := [q1,q2]:
}{}
\end{mapleinput}
\end{maplegroup}
\medskip

\noindent
We choose the ranking which compares the differentiation order of partial derivatives first and in case of equality gives priority to ${\it q1}$ over ${\it q2}$.
\medskip

\begin{maplegroup}
\begin{mapleinput}
\mapleinline{active}{1d}{ComputeRanking(ivar, dvar);
}{}
\end{mapleinput}
\end{maplegroup}
\begin{maplegroup}
\begin{mapleinput}
\mapleinline{active}{1d}{TD := DifferentialThomasDecomposition(EL, []);
}{}
\end{mapleinput}
\mapleresult
\begin{maplelatex}
\mapleinline{inert}{2d}{res := [DifferentialSystem, DifferentialSystem, DifferentialSystem]}{\[\displaystyle {\it TD}\, := \,[{\it DifferentialSystem},{\it DifferentialSystem},{\it DifferentialSystem}]\]}
\end{maplelatex}
\end{maplegroup}
\begin{maplegroup}
\begin{mapleinput}
\mapleinline{active}{1d}{PrettyPrintDifferentialSystem(TD[1]);
}{}
\end{mapleinput}
\mapleresult
\begin{maplelatex}
\mapleinline{inert}{2d}{[VectorCalculus:-`+`(q1(t), q2(t)) = 0, VectorCalculus:-`+`(VectorCalculus:-`+`(VectorCalculus:-`*`(VectorCalculus:-`*`(2, q2(t)), VectorCalculus:-diff(q2(t), t, t)), VectorCalculus:-`*`(2, VectorCalculus:-diff(q2(t), t)^2)), -1) = 0, q2(t) <> 0]}{\[\displaystyle [ \, {\it q1}(t) +{\it q2}(t) =0, \,
2\,{\it q2}(t) \left( {\frac {{\rm d}^{2}}{{\rm d}{t}^{2}}}{\it q2}(t) \right) +
2 \left( {\frac {\rm d}{{\rm d}t}}{\it q2}(t) \right) ^{2}-1=0, \, {\it q2}(t) \neq 0 \, ]
\]}
\end{maplelatex}
\end{maplegroup}
\medskip

\noindent
The first expression $q{\it1}(t)+q{\it2}(t)=0$ is a new constraint in the simple system {\tt TD[1]}.
\medskip

\begin{maplegroup}
\begin{mapleinput}
\mapleinline{active}{1d}{PrettyPrintDifferentialSystem(TD[2]);
}{}
\end{mapleinput}
\mapleresult
\begin{maplelatex}
\mapleinline{inert}{2d}{[VectorCalculus:-`+`(q1(t), VectorCalculus:-`-`(q2(t))) = 0, VectorCalculus:-`+`(VectorCalculus:-`+`(VectorCalculus:-`*`(VectorCalculus:-`*`(2, q2(t)), VectorCalculus:-diff(q2(t), t, t)), VectorCalculus:-`*`(2, VectorCalculus:-diff(q2(t), t)^2)), -1) = 0, q2(t) <> 0]}{\[ \, \displaystyle [{\it q1}(t) -{\it q2}(t) =0, \,
2\,{\it q2}(t) \left( {\frac {{\rm d}^{2}}{{\rm d}{t}^{2}}} \, {\it q2}(t) \right) +
2 \left( {\frac {\rm d}{{\rm d}t}}{\it q2}(t)  \right) ^{2}-1=0, \, {\it q2}(t) \neq 0 \, ]
\]}
\end{maplelatex}
\end{maplegroup}
\medskip

\noindent
The first expression $q{\it1}(t)-q{\it2}(t)=0$ is a new constraint in the simple system {\tt TD[2]}.
\medskip

\begin{maplegroup}
\begin{mapleinput}
\mapleinline{active}{1d}{PrettyPrintDifferentialSystem(TD[3]);
}{}
\end{mapleinput}
\mapleresult
\begin{maplelatex}
\mapleinline{inert}{2d}{[q1(t) = 0, q2(t) = 0]}{
\[
\displaystyle [ \, {\it q1}(t) =0, \, {\it q2}(t) =0 \, ]
\]}
\end{maplelatex}
\end{maplegroup}
\medskip

\noindent
These constraints are complementary to the first two constraints.
\end{example}

\begin{example}
The Landau-Lifshitz-Gilbert equations describe a unit magnetization vector
\begin{align*}
  m := \left[ \begin {array}{c} u(t) \\ v(t) \\ w(t) \end {array} \right]
\end{align*}
\mapleresult
\begin{mapleinput}
\mapleinline{active}{1d}{m := <u[0],v[0],w[0]>:}{}
\end{mapleinput}\medskip\noindent
with three dependent variables $u$, $v$, and $w$ as entries.
By computing a Thomas decomposition, we reproduce in an automated way
results from \cite{muller2007hopf}.
We assume a constant external magnetic field
\begin{mapleinput}\medskip
\mapleinline{active}{1d}{h_eff := <0,0,h3-lambda*m[3]>:}{}
\end{mapleinput}\medskip\noindent
in the direction of the $w$-axis, a self interaction of the magnetic vector in the same direction for $\lambda\in\{-1, 0, 1\}$, and we assume an additional spin torque term 
\begin{mapleinput}\medskip
\mapleinline{active}{1d}{j := <0,0,j3>:}{}
\end{mapleinput}\medskip\noindent
aligned in direction to the $w$-axis, hence counteracting the damping.

Under these assumptions the Landau-Lifshitz-Gilbert equations are given by the equations
\begin{align*}
  \left({\alpha}^{2}+1 \right) m_{{t}}
  = \alpha m \times ( m \times h_{\it eff})
  - \alpha m \times j
  - m\times h_{\it eff}
  - m\times (m \times j)
\end{align*}

\begin{mapleinput}\medskip
\mapleinline{active}{1d}{LLG := (alpha^2+1)*<u[1],v[1],w[1]>
     -alpha*CrossProduct(m,CrossProduct(m,h_eff))
     +alpha*CrossProduct(m,j)
     +CrossProduct(m,h_eff)
     +CrossProduct(m,CrossProduct(m,j)):}{}
\end{mapleinput}\medskip\noindent
where $\alpha$ is a positive real number indicating the strength of the damping. 

The alignedness assumption implies symmetry w.r.t.\ rotations around the $w$-axis.
Hence, all periodic solutions must be parallel to the equator, which implies $\frac{\operatorname{d}}{\operatorname{d}\!t}w(t)=0$.
We furthermore encode the unit length of the magnetization vector and model the parameters by differential indeterminates with derivative zero.

\begin{mapleinput}\medskip
\mapleinline{active}{1d}{LLG := [op(convert(LLG,list)),w[1],Diff2JetList(u(t)^2+v(t)^2+w(t)^2-1),
\quad alpha[1],j3[1],h3[1],lambda[1]]:}{}
\end{mapleinput}\medskip\noindent

We use the inequation $\alpha\neq 0$ to model $0<\alpha$.
To remove superfluous complex solutions, add the inequation $\alpha^2+1\neq 0$.
Furthermore, assume that the external field and the spin torque are non-degenerate, i.e., that $h_{3}$ and $j_{3}$ are non-zero.

\begin{mapleinput}\medskip
\mapleinline{active}{1d}{ineq := [alpha[0],alpha[0]^2+1,h3[0],j3[0]]:}{}
\end{mapleinput}\medskip

A Thomas decomposition of this system consists of 24 simple systems:
\begin{mapleinput}\medskip
\mapleinline{active}{1d}{ComputeRanking([t],[u,v,w,alpha,j3,h3,lambda]);
res := DifferentialThomasDecomposition(LLG,ineq):
nops(res);}{}
\end{mapleinput}
\mapleresult
\begin{maplelatex}
\mapleinline{inert}{2d}{}{\[24\]}
\end{maplelatex}\medskip\noindent
However, many of these systems contain only non-real (complex) solutions.
Remove the systems that imply one of the following four pairs of constraints:
\begin{align*}
  u(t)^2+v(t)^2 &=0 \neq v(t) \\
  j_{3}^2 + (h_{3}-\lambda)^2 &= 0 \neq h_{3}-\lambda \\
  j_{3}^2+ (h_{3}+\lambda)^2 &=0 \neq h_{3}+\lambda \\
  j_{3}^2+4\lambda^2 &=0 \neq \lambda 
\end{align*}

\begin{mapleinput}\medskip
\mapleinline{active}{1d}{l := [[u[0]^2+v[0]^2,v[0]],
\quad [j3[0]^2+(h3[0]-lambda[0])^2,h3[0]-lambda[0]],
\quad [j3[0]^2+(h3[0]+lambda[0])^2,h3[0]+lambda[0]],
\quad [j3[0]^2+4*lambda[0]^2,lambda[0]]]:
for c in l do
\quad   res := remove(a->DifferentialSystemReduce(a,c[1])=0 
\quad \quad     and c[2] in DifferentialSystemInequations(a), res):
od:}{}
\end{mapleinput}

\begin{mapleinput}
\mapleinline{active}{1d}{nops(res);}{}
\end{mapleinput}
\mapleresult
\begin{maplelatex}
\mapleinline{inert}{2d}{}{\[6\]}
\end{maplelatex}\medskip

The first system describes the generic case.
It results from incorporating some inequations and transforming the equations. 
Hence, the generically correct information can be read off this system.
First we observe the equation

\begin{mapleinput}\medskip
\mapleinline{active}{1d}{PrettyPrintDifferentialSystem(res[1])[3];
GenericEquation := Diff2JetList(lhs(%)):
}{}
\end{mapleinput}
\mapleresult
\begin{maplelatex}
\mapleinline{inert}{2d}{-alpha(t)*lambda(t)*w(t)+alpha(t)*h3(t)-j3(t) = 0}{\[\displaystyle -\alpha \left( t \right) \lambda \left( t \right) w \left( t \right) +\alpha \left( t \right) {\it h3} \left( t \right) -{\it j3} \left( t \right) \\
\mbox{}=0\]}
\end{maplelatex}\medskip\noindent
This equation holds whenenever $w(t)$ is not the constant function $1$ or $-1$:
\begin{mapleinput}\medskip
\mapleinline{active}{1d}{map(b->DifferentialSystemReduce(b,w),
  select(a->0<>DifferentialSystemReduce(a,GenericEquation),res));}{}
\end{mapleinput}
\mapleresult
\begin{maplelatex}
\mapleinline{inert}{2d}{}{\[[-1,1]\]}
\end{maplelatex}\medskip

We describe the real solutions
\begin{mapleinput}\medskip
\mapleinline{active}{1d}{with(RealDomain):}{}
\end{mapleinput}\medskip\noindent
of the first system and make some assumptions about signs of trigonometric functions to get a simplified form.
Making the opposite assumption only changes signs in the solution set.

\begin{mapleinput}\medskip
\mapleinline{active}{1d}{sol := MyPDSolve(res[1]):}{}
\end{mapleinput}

\begin{mapleinput}
\mapleinline{active}{1d}{l := [subs(sol,alpha(t))=alpha,subs(sol,lambda(t))=lambda,
\quad subs(sol,j3(t))=j3,subs(sol,h3(t))=h3]:}{}
\end{mapleinput}

\begin{mapleinput}
\mapleinline{active}{1d}{u(t)=simplify(factor(subs(l,subs(sol,u(t)))),trig) assuming alpha>0:
simplify(%) assuming cos(j3*signum(lambda)*(-t+_C5)/alpha)>0:
subs([-j3+alpha*h3+alpha*lambda=c1,-j3+alpha*h3-alpha*lambda=c2],%):
simplify(subs([c2=c3/c1],%)):
subs(c3=(-j3+alpha*h3+alpha*lambda)*(-j3+alpha*h3-alpha*lambda),%):
subs(signum(lambda)=lambda,%);}{}
\end{mapleinput}
\mapleresult
\begin{maplelatex}
\mapleinline{inert}{2d}{}{\[\displaystyle u(t) =-\frac{((h3-\lambda)\cdot\alpha-j3)((h3+\lambda)\cdot\alpha-j3)\cos\left({\frac{j3 \lambda({\it \_C5}-t)}{\alpha}}\right)}{\alpha|\lambda|\sqrt{-((h3-\lambda)\cdot\alpha-j3)((h3+\lambda)\cdot\alpha-j3)}}\]}
\end{maplelatex}

\begin{mapleinput}
\mapleinline{active}{1d}{v(t)=simplify(factor(subs(l,subs(sol,v(t)))),trig) assuming alpha>0:
simplify(%) assuming cos(j3*signum(lambda)*(-t+_C5)/alpha)>0:
subs(signum(lambda)=lambda,%);
}{}
\end{mapleinput}
\mapleresult
\begin{maplelatex}
\mapleinline{inert}{2d}{}{\[\displaystyle v(t) 
=- \sqrt{- ((h3-\lambda)\cdot\alpha-j3)  ((h3+\lambda)\cdot\alpha-j3) }
\sin \left( {\frac {{\it j3}\,\lambda\, \left( {\it \_C5}-t \right) }{\alpha}} \right)
\left(\alpha\cdot\lambda\right)^{-1}\]}
\end{maplelatex}

\begin{mapleinput}
\mapleinline{active}{1d}{w(t)=subs(l,subs(sol,w(t)));}{}
\end{mapleinput}
\mapleresult
\begin{maplelatex}
\mapleinline{inert}{2d}{}{\[\displaystyle w(t) =\frac{\alpha h3-j3}{\alpha\lambda}\]}
\end{maplelatex}\medskip

In this case $u$ and $v$ are phase shifted, and the angular velocity is $\pm\frac{j3}{\alpha}$, as $\lambda\in\{ -1,0,1\}$. 
The solutions are real if and only if

\begin{mapleinput}\medskip
\mapleinline{active}{1d}{factor(subs([j3=j3_divided_by_alpha*alpha,lambda=1],
  -(-j3+alpha*h3-alpha*lambda)*(-j3+alpha*h3+alpha*lambda)))>0:
solve(subs(alpha=1,\%),j3_divided_by_alpha) for j3_divided_by_alpha>0:
subs(j3_divided_by_alpha=j3/alpha,\%);
}{}
\end{mapleinput}
\mapleresult
\begin{maplelatex}
\mapleinline{inert}{2d}{}{\[\displaystyle \left\{ {\frac {j3}{\alpha}}<h3+1,h3-1<{\frac {{\it j3}}{\alpha}} \right\}\]}
\end{maplelatex}\medskip\noindent
holds.
(This last computation yields the same results for $\lambda=-1$.)

The solutions in the second, third, resp.\ fourth system are the same, with the additional constraints $h_3=\lambda$, $h_3=-\lambda$, resp.\ $\lambda=0$.
The last two systems contain the constant solution at the south pole and north pole.
\end{example}

\section{Conclusion}

In this paper we presented an overview of the differential Thomas decomposition method and of an implemention in Maple. (Systems of) differential equations of polynomial type arise everywhere in physics. The differential Thomas decomposition provides a universal algorithmic tool for investigating their algebraic properties and for constructing solutions by splitting the original equation(s) into a finite set of differential subsystems with disjoint solution sets. These new systems have additional algebraic properties, e.g.\ a triangular structure, the absence of multiple solutions, they disclose hidden constraints, they generate explicit conditions on parameters, and they detecte the arbitrariness in analytic solutions. The Thomas decomposition is applicable both to a single polynomially nonlinear differential equation and to a system of such equations. Moreover, one can also add inequation(s), which removes unwanted solutions. 

We illustrated these properties of the output of differential Thomas decomposition by a number of examples (Section~5). All these examples, except the first one, arise in the context of mathematical physics.

It should be noted that for an input  nonlinear differential system with several independent and dependent variables, especially with a high degree of nonlinearity, the intermediate computations can be very tedious and/or require a large amount of computer memory. More often than not, this happens when an elimination ranking is used. 

\section{Acknowledgments}
The contribution of the first author (V.P.G.) has been partially supported by the Russian Foundation for Basic Research (grant No.16-01-00080) and the RUDN University Program (5-100).

% \section{References}


%% References


\bibliographystyle{model1a-num-names}
\bibliography{DifferentialThomasCPC}






% \end{thebibliography}

\end{document}